\documentstyle[12pt]{article}

\def\be{\begin{equation}}
\def\ee{\end{equation}}
\def\ben{$$}
\def\een{$$}
\def\ba{\begin{array}{c}}

\def\ea{\end{array}}
\def\p{\partial}
\begin{document}

\titlepage
\begin{center}
.

\vspace{2cm}

{\Large \bf Perturbed P\"{o}schl-Teller oscillators
 }\end{center}

\vspace{10mm}

\begin{center}
Miloslav Znojil

\vspace{3mm}

odd\v{e}len\'{\i} teoretick\'{e} fyziky,\\
 \'{U}stav jadern\'e
fyziky AV \v{C}R, 250 68 \v{R}e\v{z},
 Czech Republic\footnote{
norbe.tex, \today; e-mail: znojil@ujf.cas.cz}\\

\end{center}

\vspace{5mm}

\section*{Abstract}

Within the framework of Lanczos-inspired perturbation theory wave
functions and energies in the short-range potential $V(x)=
a\,\rho^2(x) + \lambda\, \rho^{4} (x)$ with $\rho (x)= {\rm
sech}\, \alpha x$ and a small coupling $\lambda$ are shown
obtainable in closed form.

\vspace{10mm}

PACS
\hspace{5mm} 03.65.Ge
\hspace{5mm} 02.30.Gp
\hspace{5mm} 02.30.Hq
\hspace{5mm}
03.65.Db

\newpage

 \noindent
P\"{o}schl-Teller \cite{Poeschl} potential $V^{(0)}(x ) =
-{\mu(\mu+1)}{\rm sech}^2 \alpha\,x$ resembles the celebrated
harmonic oscillator. Within the so called shape invariant family
\cite{Khare} these two spatially symmetric exactly solvable
potentials form a unique subset.  In perturbation theory there
seems to emerge one of the most significant differences between
them.  In contrast to an enormous interest in the various
anharmonic forces \cite{Sim} there exists virtually no analysis of
a perturbed P\"{o}schl-Teller model in the current literature.
Partially, we intend to fill the gap. This letter is concerned
with the most elementary quartic example
 \be
\left [
-\frac{d^2}{dx^2}
-\frac{\mu(\mu+1)}{\cosh^2 \alpha\,x}
+\frac{4\,\lambda}{\cosh^4 \alpha\,x}
\right ]
\psi_{(PPT)}(x)
=
-\kappa_{(PPT)}^2 \psi_{(PPT)}(x). \label{SE}
 \ee
The parity is conserved, $\psi_{(PPT)}(-x)=(-1)^p
\psi_{(PPT)}(x)$, $p = 0, 1$, and all the unperturbed $\lambda=0$
bound states are available in closed form. Unfortunately, after we
re-scale $\alpha \to 1$ for simplicity we immediately notice that
the normalizable states form a mere finite set with
$\kappa_{(PPT)}^{(0)}= \mu-2N-p=\kappa(N,p) $ and wave functions
  \be
\langle x | \psi^{(0)}_{2N+p} \rangle = \frac{\tanh^p x
}{\cosh^{\kappa(N,p)} x}\cdot \ _2F_1\left ( \mu-N+\frac{1}{2},
-N, 1+\kappa(N,p), \frac{1}{\cosh^2 x} \right )
 \label{solutions0}
 \ee
where $0 \leq 2N+p < \mu$ \cite{Fluegge}. This is the reason why
these mutually orthogonal elementary functions cannot be used as
an unperturbed basis.

A key to our present $\lambda \neq 0$ construction will lie in
the use of a non-orthogonal basis. We shall employ the ansatz
 \be
\psi_{(PPT)}(x)
 = \tanh^p x\, \sum_{n=0}^\infty \,
\frac{c_n(\lambda)}
{ \cosh^{2n+\kappa_{(PPT)}} x}
  \label{sol}
 \ee
with $\kappa_{(PPT)}=\kappa(N,p) + 2\varepsilon(\lambda)$. Its
use is inspired by the general method of Lanczos \cite{Lanczos}
and its perturbative implementations \cite{DW}.  The basis
itself is taken from the particular terminating expansions
(\ref{solutions0}) which define the unperturbed coefficients
$c_n^{(0)}$ such that $c_{N}^{(0)} \neq 0$ and $c_{0}^{(0)} = 1$
while $c_{N+1}^{(0)} =c_{N+2}^{(0)} = \ldots =0$.

As always in similar constructions one inserts eq. (\ref{sol}) in
our differential eq. (\ref{SE}). This gives the current recurrence
relations which may be written in the three-term form
 \be
\lambda\,c_{n-1}+\beta_nc_n+\alpha_{n+1}c_{n+1}=0, \ \ \ \ \ \ \
n=0, 1, \ldots\  \label{recurrenc}
 \ee
or as an infinite-dimensional linear
algebraic problem
 \be
  \left(
 \begin{array}{cccc}
 \beta_0&\alpha_1&&\\
 \lambda &   \beta_1& \alpha_2&\\
 & \lambda&  \beta_2&  \ddots \\  &&\ddots&\ddots \ea \right )
\left ( \ba c_0\\ c_1\\  c_2\\ \vdots \ea \right ) = 0 \
.\label{eqon}
 \ee
In general, unfortunately, equation (\ref{eqon}) {\em cannot} be
treated as an infinite-dimensional limit of its truncated matrix
subsystems due to their non-variational, power-series origin
\cite{review,Hautot}. Benefits brought by the tridiagonality
concern only the wave functions. Their coefficients are defined by
the closed determinantal formulae \cite{class}
 \be
 c_{n+1} \sim
  \det \left(
 \begin{array}{cccc}
 \beta_0&\alpha_1&&
\\ \lambda& \beta_1&\ddots
\\ & \ddots
&  \ddots&\lambda\alpha_n\\ &&\lambda & \beta_n \ea \right) , \ \
\ \ \ \ \ n = 0, 1, \ldots \ . \label{zazrak}
 \ee
Once we postulate, in the spirit of the current perturbation
theory,
 \be
 c_n(\lambda) = c_n^{(0)}+\lambda\,c_n^{(1)}
+\lambda^2c_n^{(2)}
   + \ldots,
    \ \ \ \ \ \ \ \ \
n = 1, 2, \ldots\ ,
   \label{666}
 \ee
we may generate the separate coefficients from eq. (\ref{zazrak})
by elementary algebra. This type of dependence of the wave
functions upon the energy which is not known resembles the
Brillouin-Wigner perturbation method able to treat the energy as
an {\em external} parameter. In the similar vein the use of the
variational energies has been recommended in the semi-analytic
power-series context, e.g., by Hautot \cite{Hautot} and Tater
\cite{Tater}. The subtle problem of the determination of energies
in the present perturbative example is to be settled in what
follows.

For the sake of definitness, perturbations with $c_n(\lambda) =
c_n^{(0)} +\lambda\,h_n(\lambda)$ will be normalized to $h_0=0$.
In order to reduce the complexity of formulae we shall only pay
attention to the ground state problem with quantum numbers $p=0$
and $N=0$. This implies that for the sufficiently small
perturbations our bound state exists at any non-negative $\mu>0$
\cite{Fluegge}. The details of transition to $p=1$ and/or to $N>0$
\cite{LMP} are not too interesting in the present context. For our
choice of $p=0$ and $N=0$ the accepted normalizations simplify the
very first recurrence relation. It reads
$\beta_0+\lambda\,\alpha_1h_1=0$, defines the coefficient
$h_1={\cal O}(1)$ ``to all orders" and suggests a change of the
notation, $\beta_0=\lambda\,\gamma_0$. As long as $\gamma_0={\cal
O}(1)$ we may recall our explicit $\beta_0 \equiv \varepsilon\,
(\varepsilon+\mu+1/2)$ and infer that
$\varepsilon=\varepsilon(\lambda)={\cal O}(\lambda)$. Using a
particularly convenient ``strength-reparametrization"
$a=(2\mu+1)/4$ we shall postulate that $\varepsilon(\lambda)=
\tau(\lambda)-a$ with some not yet known analytic function. In
terms of its expansions, say,
 \be
 \tau(\lambda)=a+b\lambda+c{\lambda}^{2}+d{\lambda}^{3}
+f{\lambda}^{4}+g{\lambda}^{5} +O\left ({\lambda}^{6} \right )
\label{energyparda}
 \ee
our new ``energy-parameters" $a,  b, \ldots$ enter the matrix
elements $ \alpha_n=n/2-n^2-2n\tau $, $\beta_0=\tau^2-a^2$ and
$\beta_n=n^2+2n\tau+\beta_0$. For illustration we may insert our
ansatz in $ \beta_0=2\,ab\lambda+\left (2\,ac+{b}^{2}\right
){\lambda}^{2}+O \left ({\lambda}^{3}\right )$ giving
$\gamma_0=2\,ab+\left (2\,ac+{b}^{2}\right )\lambda+O \left
({\lambda}^{2}\right )$ etc.

Wave functions have already been specified by eq. (\ref{zazrak}).
Apparently, there are no conditions imposed upon the energies.
This is a paradox which we are going to explain now. In the first
step we insert the coefficient $h_1=-\gamma_0/\alpha_1$ in the
first, second and third row of recurrences (\ref{recurrenc}) or
(\ref{eqon}). The former relation becomes an identity but the next
one preserves a genuine three-component character,
 \be
2\,(2\varepsilon+\mu+2)\,h_2
=1+(\varepsilon+1)(\varepsilon+\mu+3/2)\,h_1.
 \label{6}
  \ee
Up to a tiny perturbation the infinite rest with $n = 1, 2,
\ldots$ has just a two-term form
 \be
(n+2)(2\varepsilon+n+\mu+2)\,h_{n+2}
=
(\varepsilon+n+1)(\varepsilon+n+\mu+3/2)\,h_{n+1} +\lambda\,h_n
\label{3}
 \ee
giving immediately the asymptotic estimate valid for all
$|\lambda| \ll 1$,
 \be
h_n
=h_n(\lambda)
 \sim
\frac{\Gamma(\varepsilon+n)\Gamma(\varepsilon+n+\mu+1/2)}
{n!\Gamma(2\varepsilon+n+\mu+1)}
\sim
 n^{-3/2}, \ \ \ \ \ n \gg 1.
 \label{gamas}
 \ee
This implies the convergence of $\psi_{(PPT)}(x)$, up to the
central $x= 0$, rigorously.

The boundary of the circle of convergence coincides with the
centre of the spatial symmetry. Hence, for our particular
even-parity choice of $p=0$ the first derivative of the wave
function must have a nodal zero there,
 \be
 \p_x\psi_{(PPT)}(x)\sim \sinh x\ \sum_{n=0}^\infty
 \frac{ (2n + \kappa)\,h_n}{\cosh^{2n+\kappa+1}x}
\to 0, \ \ \ \ x \to 0.
 \label{expr}
 \ee
As long as $n \,h_n \sim n^{-1/2}$ the infinite sum itself is
divergent in the origin. The whole expression (\ref{expr})
exhibits an $0 \times \infty$ indeterminacy there. The
conservation of parity must be imposed ``by brute force"
\cite{LMP}. Vice versa, the strict validity of the boundary
condition $ \p_x\psi_{(PPT)}(0)=0$ represents precisely the
``seemingly lost" quantization condition.

In the nearest vicinity of $x=0$ the finite but very large value
of the sum in eq. (\ref{expr}) is equal to a positive real number
multiplied by the coefficient $h_2$. This observation follows from
eq. (\ref{3}) which gives, step-by-step, $h_3 (0) =
2(\mu+5/2)/(3\mu+9)\cdot h_2 (0) $, $h_4 (0) = 3(\mu+7/2)/(4\mu
+16)\cdot h_3 (0)$ etc. In the interval
$\varepsilon>-(\mu+3)/2$ all the multiplication factors remain
positive. This guarantees that the parity is conserved, in a
semi-infinite interval of $\lambda$, if and only if the
coefficient $h_2(\lambda)$ vanishes in the leading order
approximation.

This is one of our most important observations. The required
leading-order change of sign of $h_2(\lambda)$ (i.e., relation $
h_2^{(1)}=0$) may be interpreted as the consequence of the
Sturm-Liouville oscillation theorems \cite{Hille}. In their light,
for an increasing or decreasing leading-order energy
$E=-\kappa^2=-[2a-1/2+{\cal O}(\lambda)]^2$ the nodes of $
\p_x\psi_{(PPT)}(x)$ would {\rm smoothly} move along the real axis
of $x$. In this sense the condition $h_2^{(1)}=0$ of coincidence
of one of the nodes with the origin reflects the zero-order
physical interpretation of the first coefficient $a$ in our
expansion (\ref{energyparda}) and forces us to postulate,
consequently, the disappearance of all the subsequent
leading-order coefficients,
 \be
h_2^{(1)}=h_3^{(1)}=\ldots =0. \label{zeros}
 \ee
Before writing it down as an explicit algebra let us first move up
to the next order approximation. Due to the danger of a
re-introduction of the asymmetry (or, in the other words, of a
spike-shaped discontinuity in $\psi_{(PPT)}(x)$ at $x=0$
\cite{LMP}) in the higher orders of $\lambda$ we just have to
repeat the previous argumentation. In the second order we must
first notice that the vanishing condition (\ref{zeros}) changed
also our recurrences (\ref{eqon}). The role of the ``last genuine
three-term" relation (without a guarantee of a sign-preservation)
moves from $n=1$ to $n=2$ in eq. (\ref{recurrenc}). As long as our
estimate (\ref{gamas}) remains valid within its own error bound
$1+{\cal O}(1/n)$ the role of an overall sign-determining factor
$\lambda\,h_2(\lambda)={\cal O}(\lambda^2)$ is taken over by the
new norm $\lambda\,h_3(\lambda))={\cal O}(\lambda^2)$. {\it
Mutatis mutandis} we get the second-order condition $
h_3^{(2)}=h_4^{(2)}=\ldots =0$ and, in general,
 \be
c_n(\lambda) = \lambda\, h_n(\lambda)=
\lambda^{n}h_n^{(n)}+\lambda^{n+1}h_n^{(n+1)}
   + \ldots
  \
,
   \ \ \ \ \ \ \ \ n = 1, 2, \ldots\ .
    \ee
This is a perturbative generalization of the standard termination
rules. It modifies our above naive expectations and incorporates
correctly the physical definite-parity requirement in our original
ansatz (\ref{sol}). In a more compact notation let us put
$c_0=f_0(\lambda)=1$ and $ \lambda\, h_n(\lambda) \equiv
\lambda^nf_n(\lambda)$ with $f_n(\lambda)={\cal O}(\lambda^0)$.
This modifies our system of equations (\ref{eqon}),
 \be
  \left(
 \begin{array}{ccccc}
 \beta_0&\lambda\alpha_1&&&\\
 \lambda&  \lambda \beta_1& \lambda^2\alpha_2&&\\
 & \lambda^2&  \lambda^2 \beta_2& \lambda^3\alpha_3&\\
& & \lambda^3&  \lambda^3 \beta_3& \ddots \\ & &&\ddots&\ddots \ea
\right ) \left ( \ba f_0\\ f_1\\f_2\\f_3\\ \vdots \ea \right ) =
0. \ \label{eqondia}
 \ee
All these equations may be solved by the closed determinantal
formula again,
 \be
 f_{n+1}(\lambda)= \frac{
\Gamma(\mu+1+2\lambda\eta)}{ \Gamma(\mu+n+2+2\lambda\eta)} \cdot
\left [ {\lambda^{-(n+1)(n+2)/2}}{{\cal M}_n(\lambda)}
 \right ],
 \label{cons}
 \ee
  \ben
{\cal M}_n(\lambda)=
  \det \left(
 \begin{array}{ccccc}
 \beta_0(z)&\lambda\alpha_1(z)&&&
\\ \lambda&\lambda \beta_1(z)& \lambda^2\alpha_2(z)&&
\\ & \lambda^2& \ddots&\ddots &
\\ && \ddots
&  \lambda^{n-1}\beta_{n-1}&\lambda^n\alpha_n\\ &&&\lambda^n &
\lambda^n\beta_n \ea \right)\ .
 \een
It pre-factorizes the wave function coefficients in a certain less
usual way. We may re-shuffle it slightly once more, in the more
traditional Pad\'e-approximation spirit, with $f_0=\gamma_{-1}=1$
and
 \be
 f_j=\frac{\gamma_{j-1}}{(
 -\alpha_1)(-\alpha_2)\ldots (-\alpha_j)} \,, \ \ \ \
\ \ \ \
 \, j = 1, 2, \ldots\ .
  \ee
This representation of the wave function coefficients is related
to our final, ``optimal" recurrences
 \be
  \left(
 \begin{array}{ccccc}
 \beta_0&-\lambda&&\\
 -\alpha_1&  \beta_1&- \lambda&\\
 & -\alpha_2&   \beta_2&-\lambda  & \\  &&\ddots&\ddots &\ddots
 \ea \right )
 \left (
  \ba 1 \\ \gamma_0\\ \gamma_1\\
  \vdots \ea \right ) = 0 \ .
\label{eqontria}
 \ee
The proof of their equivalence to our previous equations is
trivial. One has only to verify that the new notation is not
inconsistent since the first row of eq. (\ref{eqontria}) just
reproduces our old leading-order ``change of the notation"
$\beta_0=\lambda\,\gamma_0$.

We are now prepared to compute the unknown auxiliary variables $b,
c, \ldots$ and, in effect, the perturbed energies. Firstly we get
rid of the ``already known" $\gamma_0$ in eq. (\ref{eqontria}).
This only means that we omit the first line and abbreviate
$\hat{A}_2=\alpha_2\gamma_0$ and $ \hat{B}_1= \beta_1
\gamma_0-\alpha_1 $ in the remaining equations
 \be
  \left(
 \begin{array}{ccccc}
   \hat{B}_1&- \lambda&&&\\
  -\hat{A}_2&   \beta_2&-\lambda  && \\
  & -{\alpha}_3&   \beta_3&-\lambda  & \\
 &  &\ddots&\ddots &\ddots
 \ea \right )
 \left (
  \ba 1 \\ \gamma_1\\ \gamma_2\\
  \vdots \ea \right ) = 0 \ .
\label{ontria}
 \ee
Now we may parallel the above study where, retrospectively, $
\hat{B}_0=\beta_0$. In the other words we only have to require
that $ \hat{B}_1 \equiv \lambda\,\gamma_1$ vanishes in the
unperturbed limit (cf. the first line in eq. (\ref{ontria})). In
contrast to its trivial $j=0$ predecessor the new $j=1$ situation
imposes a constraint upon $ \hat{B}_1= 2a+{1}/{2}+2ab(1+2a)+{\cal
O}(\lambda)$. Its zero-order component must be zero. This
determines the physical value of the first moment $b=b(a)=
\tau^{(1)}$,
 \be
\tau^{(1)}=-{\frac {1+4\,a}{4\,a\,\left (1+2\,a\right )}}.
 \label{porvy}
 \ee
As expected on variational grounds this result means a positive
first order change in our energy $E=-\kappa^2$ where $\kappa
=\kappa(\lambda) =[4\tau(\lambda)+2\mu-1]/4$.

The whole $j \to j+1$ procedure can be, obviously, iterated. We
arrive at the equations
 \be
  \left(
 \begin{array}{ccccc}
   \hat{B}_j&- \lambda&&&\\
  -\hat{A}_{j+1}&   \beta_{j+1}&-\lambda  && \\
  & -{\alpha}_{j+2}&   \beta_{j+2}&-\lambda  & \\
 &  &\ddots&\ddots &\ddots
 \ea \right )
 \left (
  \ba 1 \\ \gamma_{j}\\ \gamma_{j+1}\\
  \vdots \ea \right ) = 0 \
\label{gentria}
 \ee
with the elements $ \hat{A}_{j+1}= \alpha_{j+1}\gamma_{j-1}$ and $
\hat{B}_j= \beta_j \gamma_{j-1}-\alpha_j\gamma_{j-2}$. For any
$j$, as we have already shown, the first-line rule $ \hat{B}_j=
\lambda \gamma_{j}= {\cal O}(\lambda)$ determines the value of the
moment $\tau^{(j)}$. Beyond the formal choice of $j=0$ (confirming
the consistency of $a= \tau^{(0)}=\mu/2+1/4$) and of $j=1$ (giving
$b=\tau^{(1)}$) we must perform symbolic manipulations which
become a task for the computer. Without its assistance one easily
obtains just the next and quite compact $j=2$ formula
$$c=\tau^{(2)}=-1/32\,{\frac {\left (1+4\,a\right )\left
(8\,{a}^{2}+8\,{a}^{3}+3\,a+ 1\right )}{\left (1+2\,a\right
)^{3}{a}^{3}\left (1+a\right )}}. $$ {\em With} the assistance of
the MAPLE language \cite{Maple} the evaluation of the next
corrections remains entirely straightforward giving $d$ or
$$\tau^{(3)}= -{\frac {\left (1+4\,a\right ) {\cal D}_{7}
}{{{128}}\,{a}^{5}\left (1+2\,a\right )^{5}\left (1+a\right )^{2}
\left (3+2\,a\right )}} $$ with $$ {\cal
D}_{7}=3+26\,a+97\,{a}^{2}+168\,{a}^{3}+ 196\,{a}^{4}+
288\,{a}^{5}+320\,{a}^{6}+128\,{a}^{7} $$ or $f$ or rather
 $$\tau^{(4)}= -{\frac {\left
(1+4\,a\right ){\cal D}_{12} }{{{2048}}\,{a}^{7}\left
(1+2\,a\right )^{7 }\left (1+a\right )^{3}\left (3+2\,a\right
)^{2}\left (2+a\right )}} $$ with $$ {\cal D}_{12}=90+
1335\,a+8815\,
{a}^{2}+32715\,{a}^{3}+69135\,{a}^{4}+54250\,{a}^{5}$$ $$
-106568\,{a}^{6}-
340152\,{a}^{7}-378096\,{a}^{8}-165184\,{a}^{9}+22272\,{a}^{10}+43008
\,{a}^{11}+10240\,{a}^{12}$$ etc. These results inspire the
general ansatz
 \ben
\tau^{(K)}= -{\frac {{{2^{-M(K)}}}\,\left (1+4\,a\right ){\cal
D}_{L} }{{a}^{2K-1}\left (1+2\,a\right )^{2K-1 }\left
(a+1\right )^{K-1} \left (a+3/2\right )^{K-2}
 \left (a+2\right )^{K-3}\ldots \left
(a+K/2\right )
 }}
 \een
re-confirmed by the explicit evaluation of the next-order
correction $g$,
 $$\tau^{(5)}=-{\frac {\left
(1+4\,a\right ){\cal D}_{18} }{{{8192}}\,{a}^{9} \left
(1+2\,a\right )^{9}\left (1+a\right )^{4}\left (3+2\,a\right )^{
3}\left (2+a\right )^{2}\left (5+2\,a\right )}} $$ with the rather
complicated
 $${\cal
D}_{18} =3780+80892\,a+794817\,{a}^{2}+\ldots+458752\,{a}^{18}. $$
This did not disprove the obtrusive hypothesis that $ {\cal D}_{L}
$ are polynomials in $a$ with integer coefficients and with a
growing degree $L=L(K)=(K+1)(K+2)/2-3$.

We may summarize our approach to eq. (\ref{SE}) as a scheme which
can be easily generalized. Its use of a suitable non-orthogonal
basis may find applications in different settings. In a way
paralleling the Lanczos method (with a strong numerical flavor)
the idea itself has already been tested on several ``less
solvable" semi-numerical examples \cite{DW,LMP}. In contrast to
them our present construction seems to exhibit a much closer
similarity to the current perturbation constructions of anharmonic
oscillators. Even in this comparison, our new example seems to
hide several pleasant surprises (e.g., a possible non-zero radius
$\lambda_{max}$ of convergence) which could make it worth of a
further study.

\subsection*{Acknowledgement}
Our thorough clarification of the ``subtle problem of the
determination of energies" has been encouraged by an anonymous
referee. The work was supported by the GA AS {C}R, grant Nr.
A1048004.

 \end{document}